\providecommand{\pgfsyspdfmark}[3]{}
\documentclass{iau} 
\usepackage{natbib}
\usepackage{graphicx}
\usepackage{hyperref}

\title[X-ray Polarization ] 
{X-ray Polarization at the Crossroads }

\author[Heyl, Caiazzo \& Gonz\'alez-Caniulef]   
{Jeremy Heyl$^1$, Ilaria Caiazzo$^2$, Denis Gonz\'alez-Caniulef$^1$
}

\affiliation{$^1$Department of Physics and Astronomy, University of British Columbia,\\ Vancouver BC V6T 1Z1, Canada\\
$^2$ TAPIR, Walter Burke Institute for Theoretical Physics, Mail Code 350-17, Caltech, Pasadena, CA 91125, USA\\}

\pubyear{2022}
\volume{363}  
\jname{Neutron Star Astrophysics at the Crossroads:
Magnetars and the Multimessenger Revolution}
\editors{E. Troja \& M. Baring, eds.}
\begin{document}

\maketitle

\begin{abstract}
We argue that measurements of X-ray polarization using the recently launched Imaging X-ray Polarimetry Explorer will answer many open questions about magnetars in particular the physical state of their surfaces, whether vacuum birefringence exists, and the nature of the hard X-ray emission from these objects.  We outline the capabilities of the instrument, specific models and the results of simulations for the magnetar 4U~0142+61.
\end{abstract}

\firstsection 
\section{Introduction}

We stand at the threshold of a new era in high-energy astrophysics.  In December 2021, the Imaging X-ray Polarimetry Explorer (IXPE) launched from Cape Canaveral and during February 2022 it will provide the first polarimetric observations of a magnetar in X-rays, specifically of the AXP 4U~0142+614.  We will present an overview of what to expect from IXPE observations of the magnetar 4U~0142+614 and how these observations will inform our understanding of magnetars and neutron stars more generally.  In fact, it is our sincere hope that several of the models presented here will actually be excluded by IXPE observations in the near future.

\section{IXPE}

\begin{figure}
\centering
    \includegraphics[width=0.48\textwidth,clip,trim=2in 0 2in 0]{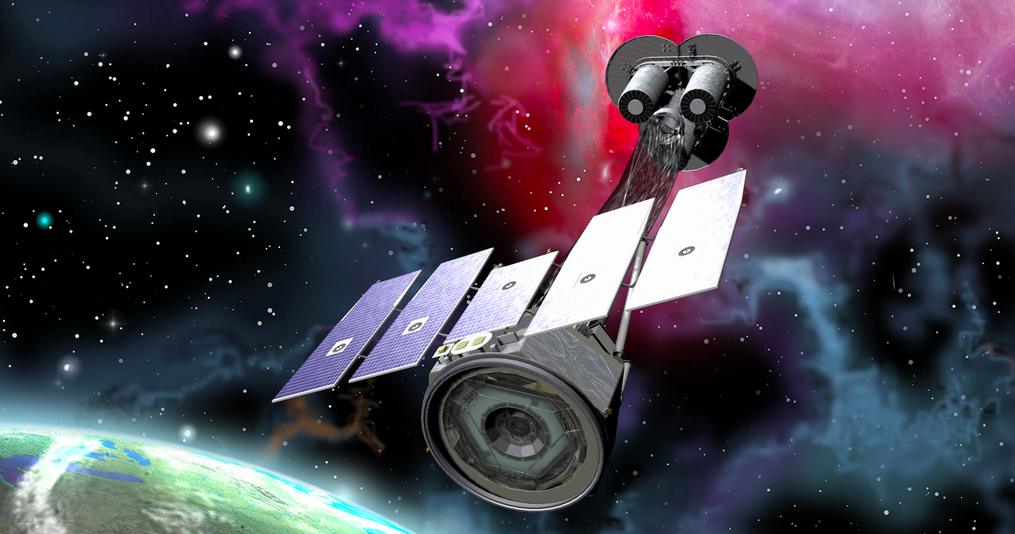}~~
        \includegraphics[width=0.48\textwidth,clip,trim=0 0 0 0]{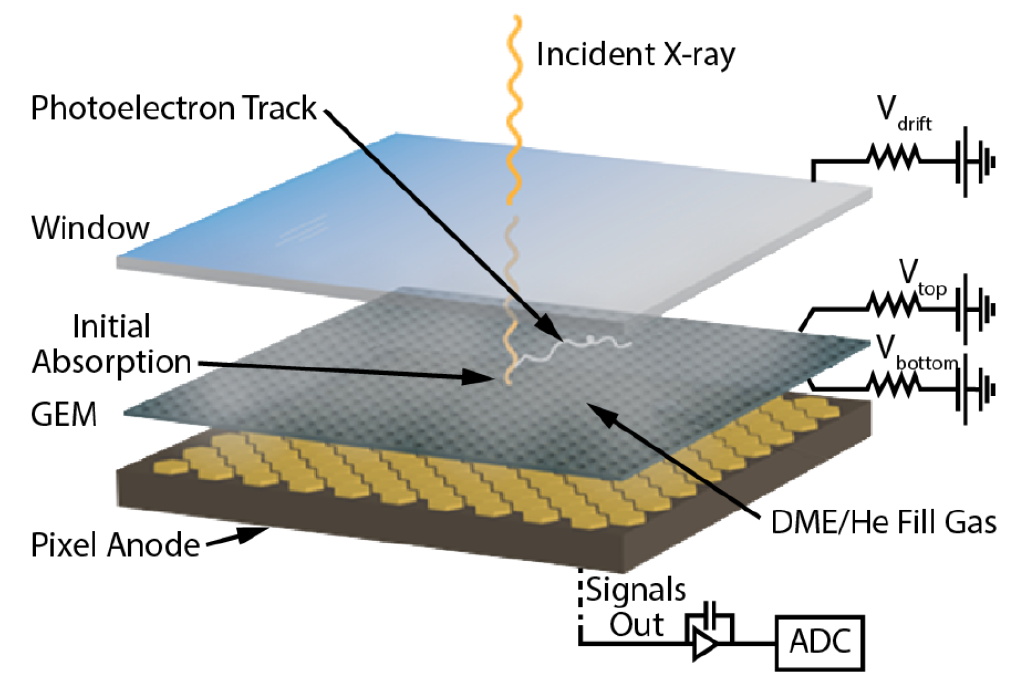}
    \caption{Left: IXPE in Orbit. Right: Gas Pixel Detector \citep[Credit: NASA;][]{2017SPIE10397E..0FS}}
    \label{fig:ixpe}
\end{figure}

NASA’s IXPE was launched in December 2021.  It is 100 times more sensitive to polarized X-ray emission than any preceding mission, and it is opening a new observational window in high-energy astrophysics. The IXPE observatory (Fig.~\ref{fig:ixpe} left) consists of three X-ray telescopes mounted in parallel, each feeding radiation into a gas-pixel detector (Fig.~\ref{fig:ixpe} right).  The X-ray photons are absorbed by atoms in the gas (dimethyl ether) and ionize an electron from the K-shell.  The ionization is typically well above the threshold of the particular atom, so the cross section decreases with increasing photon energy.  Furthermore, the outgoing electron is approximately in a well-defined momentum state; the direction of the momentum correlates with the photon electric field with a $\sin^2 \theta$ dependence.  

The magnitude of the correlation for fully linearly polarized radiation is known as the modulation factor.  Occasionally, the atom will emit an additional low-energy Auger electron.  By modelling the high-energy photo-electron track \citep{2017SPIE10397E..0FS}, the following properties of the incoming photon can be determined: energy, arrival time, electron direction and position on the sky.  As the energy of the photon increases, so does the momentum of the photo-electron, which results in a longer and more easily measured track, and therefore the modulation factor for IXPE increases with increasing energy from 15\% at 2~keV to 60\% at 8~keV \citep{2021arXiv211201269W}.  The richness of these data pushes the development of more sophisticated analysis techniques such as unbinned likelihood to get the most from the data \citep{2021AJ....162..134M}.   

The total effective area of the three mirror arrays is about 600~cm$^2$.  As mentioned earlier, the cross section of photoionization decreases with increasing photon energy; therefore, once the quantum efficiency of the detector is included the total effective area, it decreases to about 75~cm$^2$ at 2~keV and 8~cm$^2$ at 7~keV \citep{2021arXiv211201269W}.  Although the effective area drops quickly with increasing energy, the increase in modulation factor mitigates this for the detection of polarization.  IXPE is not sensitive to circular polarization.  The energy resolution of the detectors is typically about 20\%, the angular resolution is about 20~arcseconds, and the time resolution is one microsecond.   All told, these specifications make IXPE a factor of 100 more capable than any X-ray polarimeter in space before it.  The discovery potential is vast.

\section{QED and Magnetars}
\label{sec:QED}

Since the discovery of neutron stars and their strong magnetic fields, a key question is whether observations of neutron stars could probe the radiative corrections of quantum electrodynamics (QED) in a strong magnetic field \citep{Heis36,Weis36,Schw51}.  Most important for our discussion is how vacuum birefringence \citep{Tsai75,1997JPhA...30.6485H} can affect the polarization from neutron stars.  First came the argument of \citet{1977ApJ...215L.117N} that vacuum birefringence would generate a large phase difference between the normal modes and significant depolarization, and that ``it is extremely unlikely that polarized X-ray emission which arose from the surface of a
magnetic neutron star would be observed.''   \citet{1978SvAL....4..117G} countered the argument of \citet{1977ApJ...215L.117N} by pointing out that a large phase retardation between the normal modes is a necessary but not sufficient condition for depolarization, and in fact, if the emission consists of an incoherent superposition of normal waves, no depolarization will occur unless the properties of the birefringence vary too quickly.  Birefringence would preserve the polarization along the path of the radiation.  Furthermore, later work demonstrated that vacuum birefringence causes the direction of polarization in the X-rays for typical neutron stars to follow the direction of the magnetic field (the radiation remains in one of the normal modes) until a distance far from the surface of the neutron star, the polarization-limiting radius \citep{1999MNRAS.306..333S,2000MNRAS.311..555H}.  In particular, this QED effect makes the observed polarization many times larger than one would expect from emission coming from a large portion of the neutron star surface \citep{2002PhRvD..66b3002H}. 

\begin{figure}
\centering
\def\fsize{0.45\columnwidth}
    \includegraphics[height=\fsize,clip,trim=0.85in 0 0.3in 0]{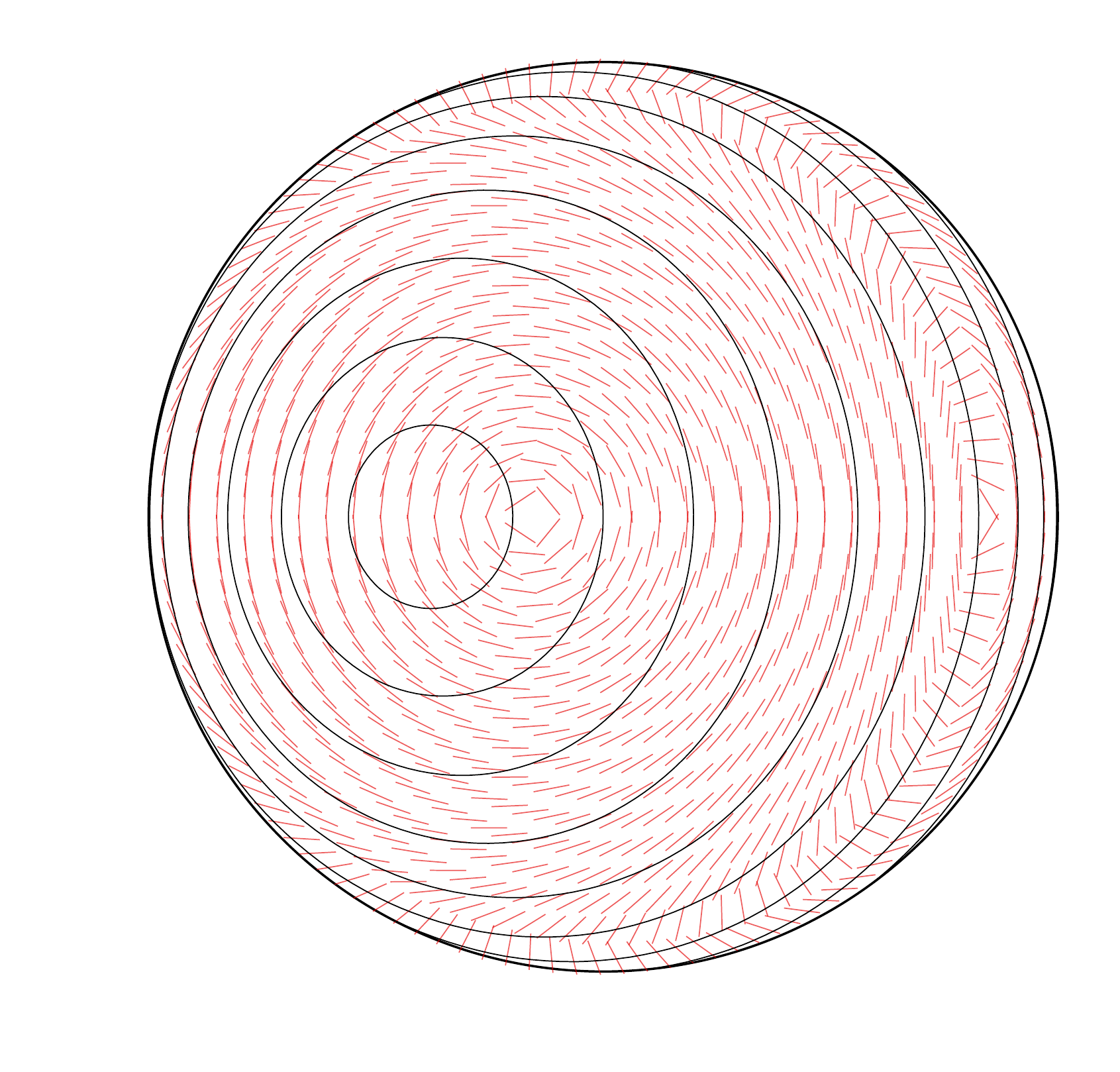}     
   \includegraphics[height=\fsize,clip,trim=0.85in 0 0.3in 0]{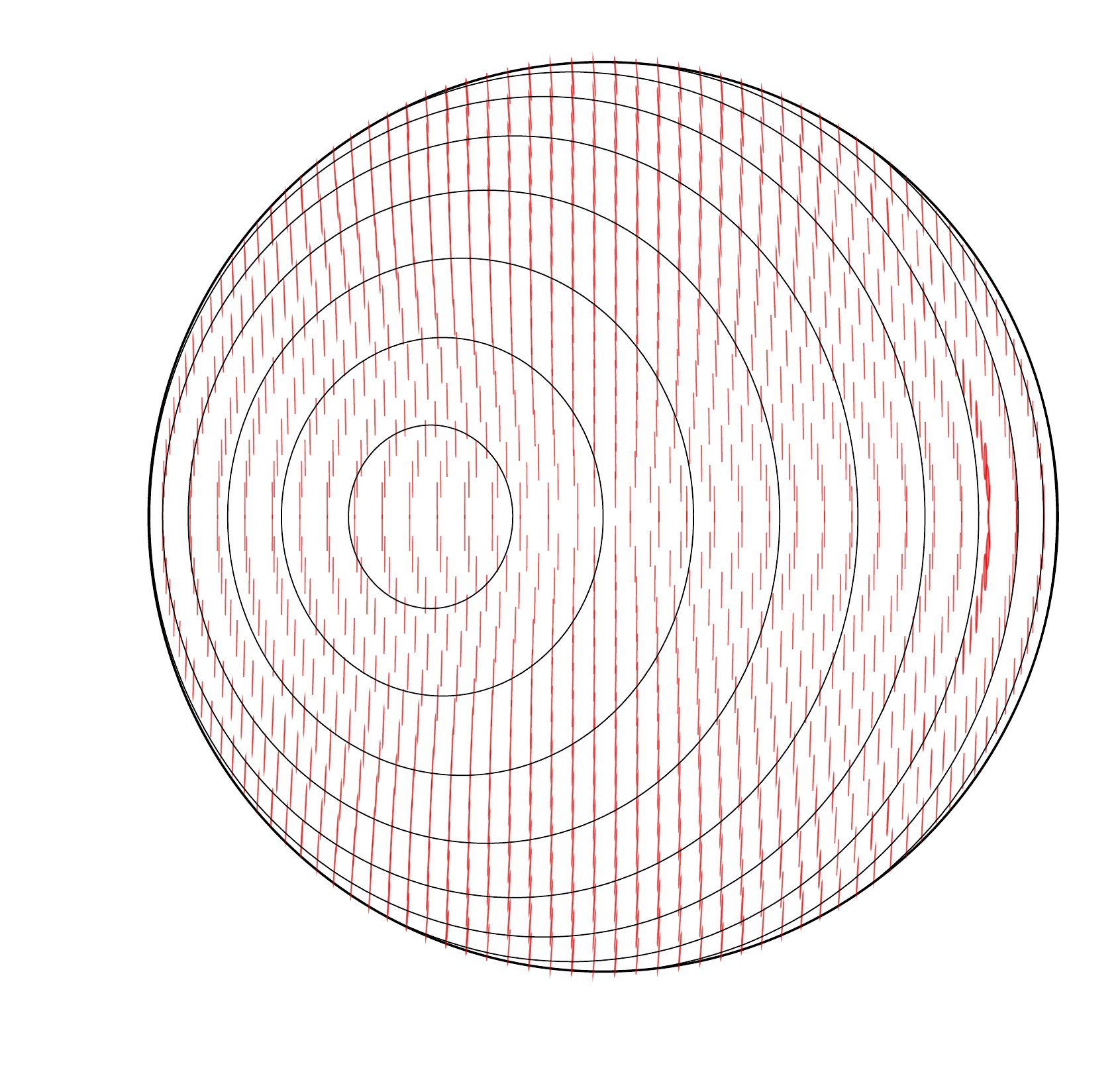}
       \includegraphics[height=\fsize]{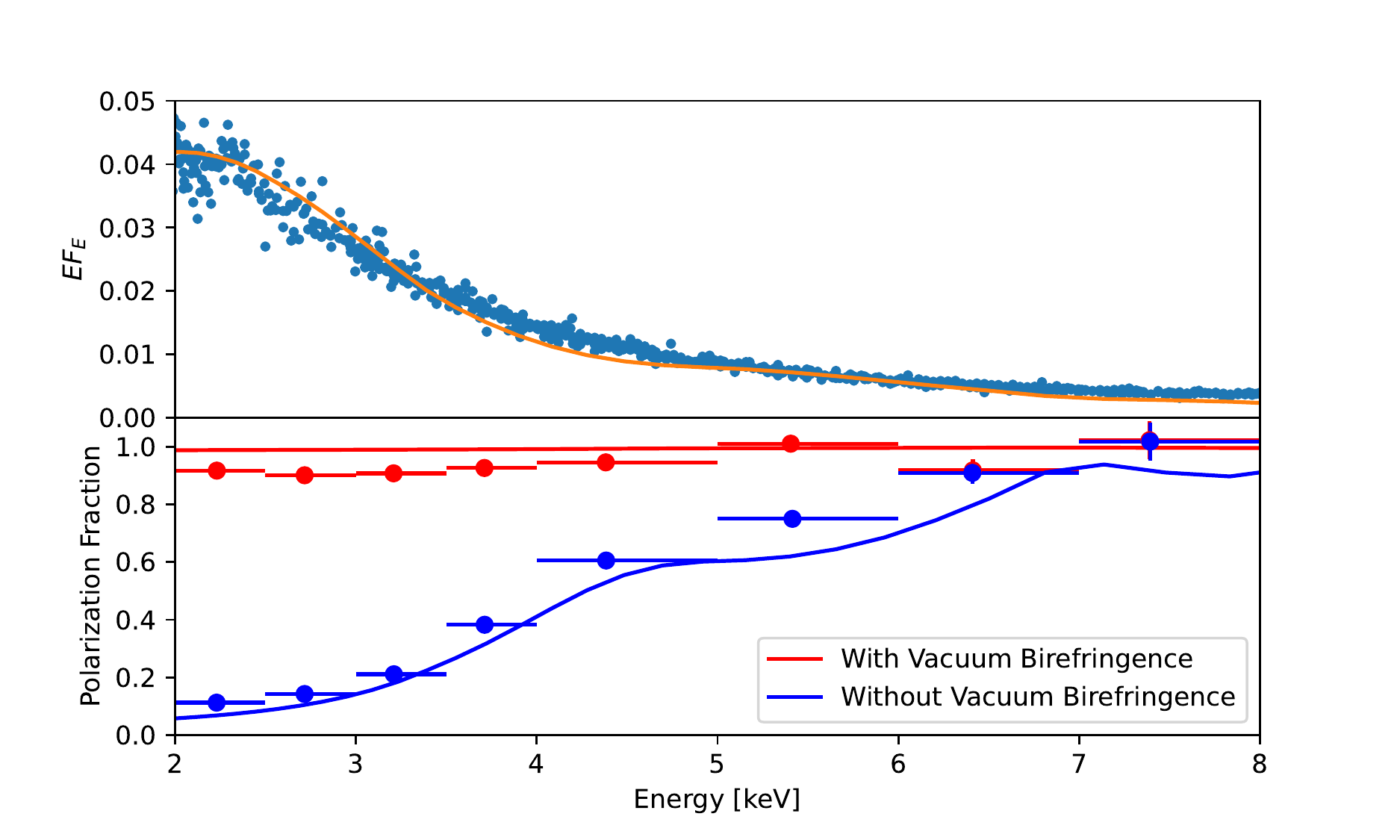} \\
    \caption{Top Left and Right: Polarisation maps from \citet{2018Galax...6...57C} without and with QED.  The curves are contours of constant magnetic latitude (separated by 15 degrees), and the lines denote the extraordinary mode.  Bottom: Predicted mean phase-resolved polarization of 4U~0142+61 and phase-averaged flux \citep{2015ApJ...808...32T} from the atmospheric models of \citet{2021arXiv211203401C} with a simulated 1~Ms observation with IXPE \citep{HeylGnedin2022}. }
   \label{fig:surfmap}
\end{figure}

The result of this effect is shown in Fig.~\ref{fig:surfmap}.  The left-hand panel shows the polarization map at the surface of the star, in particular, the direction of the polarization of the extraordinary mode.  Across the stellar surface, the direction varies significantly, so if one were to sum the total polarized radiation in this case, one would obtain a small polarized fraction.  The right panel depicts the situation with vacuum birefringence.  At a large distance from the surface of the star, the magnetic field will follow a dipole pattern; its projection in the plane of the sky will be parallel to the magnetic axis, in this case horizontal.  The polarization vectors that still follow the extraordinary mode are all vertical and the total polarization integrated over the surface of the star can be large if the polarization fraction at emission is large.

\section{Magnetar Emission Mechanisms}

Beside probing the effect of vacuum birefringence, polarimetric observations of the magnetars will determine the physical state at the photospheres of neutron stars: solid, liquid or gas. Although both gaseous and condensed surfaces can account for the soft X-ray emission from these objects, they yield dramatically different polarization signatures \citep{2021arXiv211203401C}.   Underneath the surface layers, we assume that the flux from below follows a dipole distribution \citep{1998MNRAS.300..599H}, with the possible inclusion of a small, hotter polar cap.

\subsection{Atmosphere}
\label{sec:atm}

In analogy to white dwarfs, the strong gravitational acceleration at the surface of the neutron star will causes the uppermost layers to consist of the lightest elements present.  If the neutron star has experienced even a modest amount of accretion (more than $10^{14}$~g), then the atmosphere will consist of hydrogen; otherwise, it is expected to consisted of elements in the iron group \citep[e.g.][]{2009MNRAS.400.1986H}.  Although the details of the spectral energy distribution will depend on the composition of the atmosphere (e.g. lines and the shape of the spectrum), the signatures in the polarization depend most strongly on the polarization modes of the strongly magnetized plasma that comprises the atmosphere, and therefore they only depend weakly on the detailed composition of the atmosphere.  For clarity, we will focus on the results for an atmosphere consisting of fully ionized hydrogen atoms, i.e. electrons and protons \citep{lloyd03a,lloyd03b}.

Rather than review the broad work on neutron star atmospheres in detail, we will focus on a few key ideas that influence the expected polarization more dramatically.  First, as the emission from the atmosphere is well below the cyclotron energy, $E \sim 1~\textrm{keV} \ll\ E_\textrm{cyc} \sim 1~\textrm{MeV}$, the propagation modes of the radiation in the plasma are the extraordinary mode, where the transverse electric field does not have a component along the magnetic field direction, and the ordinary mode, where the transverse electric field does have a component along the magnetic field.  Loosely speaking, the cross-section for ordinary photons to interact with the electrons in the magnetized plasma is similar to the situation where there is no magnetic field (hence the name ordinary). For the extraordinary mode, the situation is somewhat more complicated in that the longitudinal component of the electric field that can interact with the plasma electrons is very small ($\sim E^2/E_\textrm{cyc}^2$); consequently, cross sections for interactions between the extraordinary mode and the plasma are diminished by a factor of about $E^2/E_\textrm{cyc}^2$.  
\begin{figure}
\centering
\includegraphics[width=0.7\columnwidth]{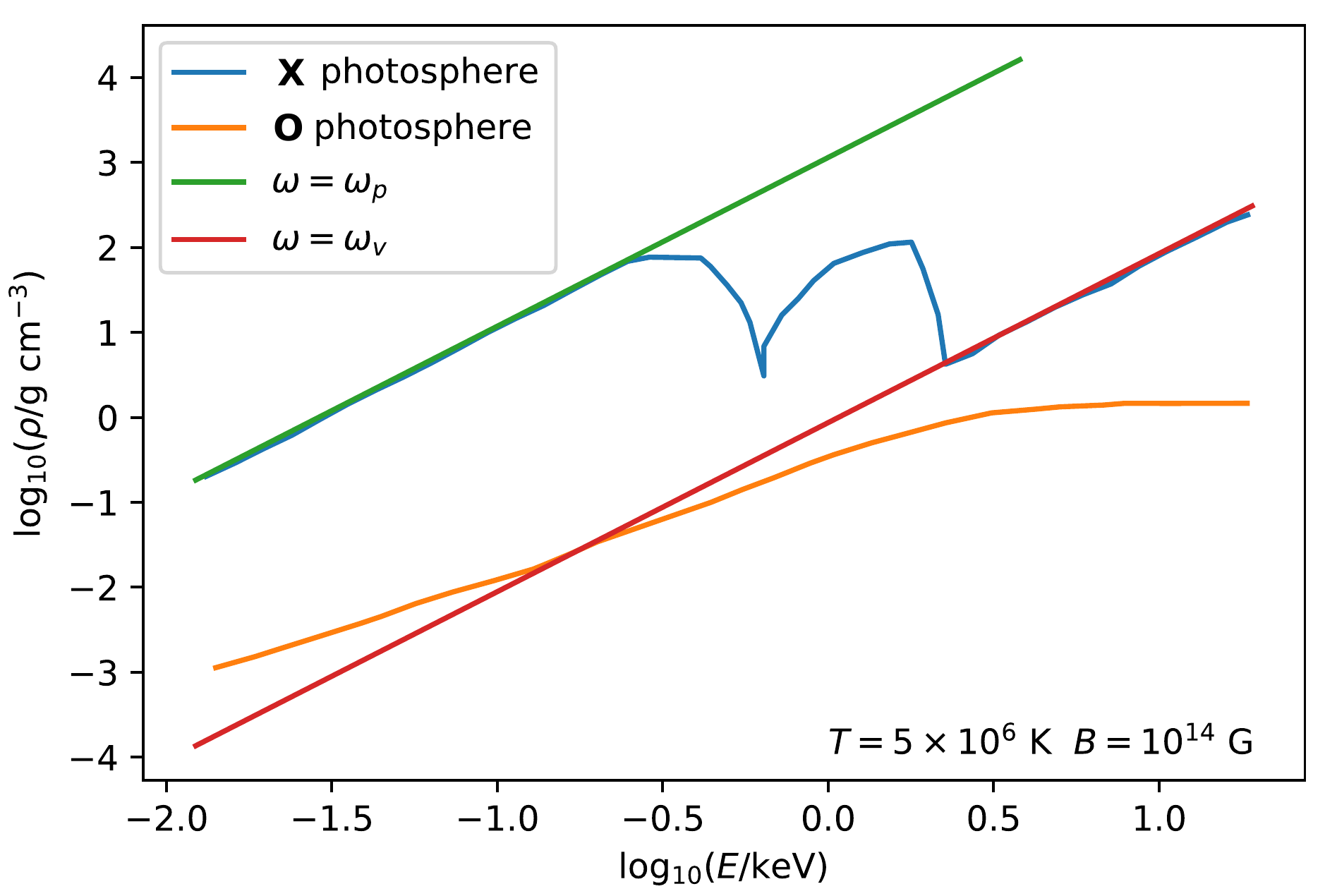}     \includegraphics[width=0.7\columnwidth,clip,trim=-5mm 0 0 0]{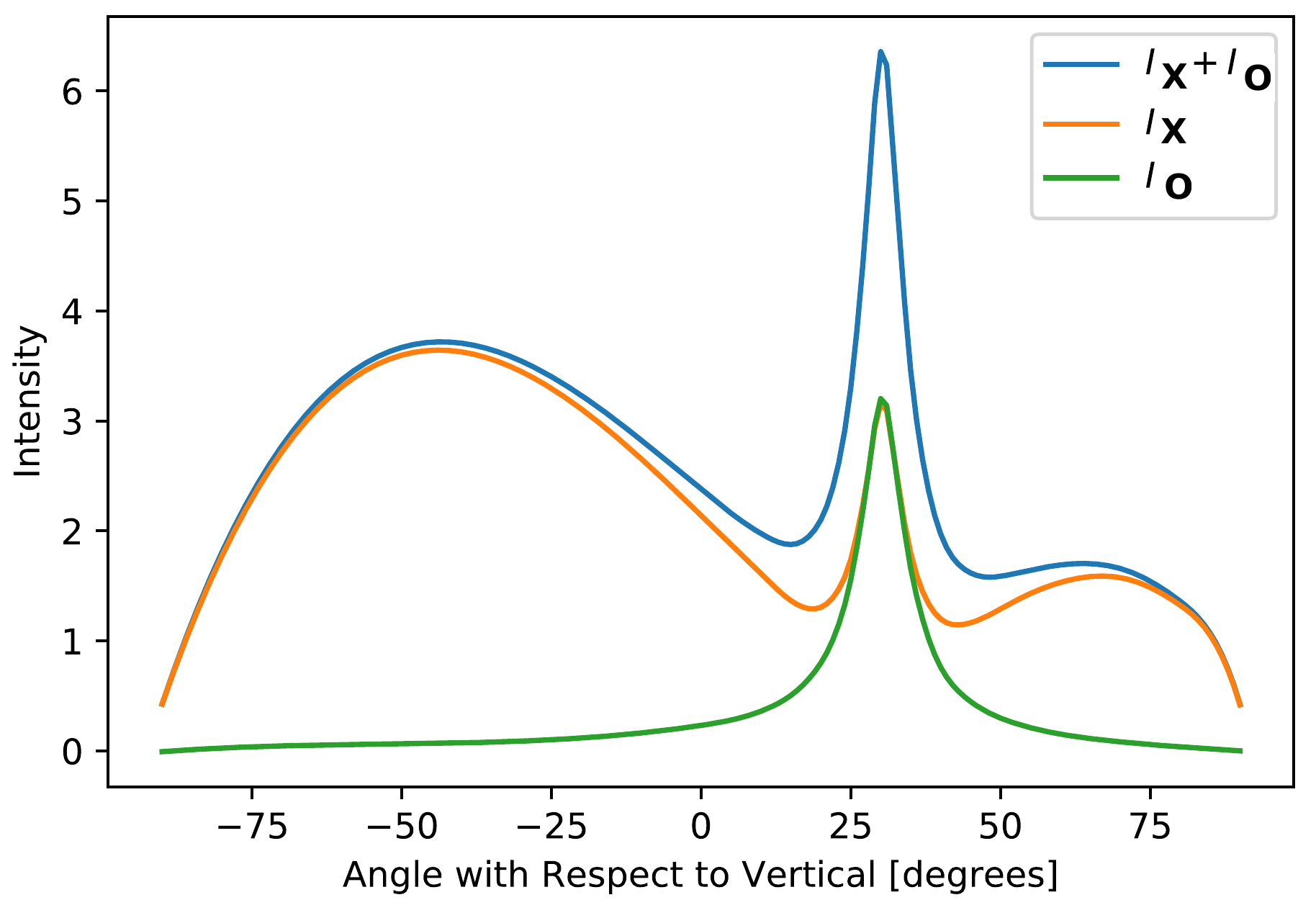} 
    \\
    \caption{Upper: Schematic of Effective Photospheres as a Function of Energy and Polarization. Lower: Beaming of Polarized Radiation with respect to the Magnetic Field \citep{2019ASSL..460..301C}.  }
    \label{fig:atmo}
\end{figure}

The upper panel of Fig.~\ref{fig:atmo} shows how this affects the propagation of the radiation of the two modes in the atmosphere.  The location of the photosphere indicates where the mean-free path of a photon is similar to the scale height of the atmosphere; from here is where the photons emerge.  The cross section for low-energy photons ($E<0.2$~keV) in the extraordinary mode to interact with the plasma is so low that their photosphere is determined by the plasma frequency, the collective response of the electrons in the plasma rather than interactions between photons and individual electrons.  From 0.3~keV to 3~keV, the interaction for ordinary and extraordinary photons is dominated by free-free absorption.  For the extraordinary mode, the cross-section is approximately constant with energy so the photosphere lies at constant density.  For the ordinary mode, the cross section decreases with increasing photon energy, so the photosphere is deeper for higher energy photons until the cross-section reaches the Thompson cross-section and the photosphere lies at constant density above this energy. 

In the upper panel of Fig.~\ref{fig:atmo} we see that the situation for the extraordinary mode can be more complicated.  The feature at just below 1~keV is the proton cyclotron resonance, which causes the extraordinary mode to interact strongly so that near the resonance, the cross-section for interaction is larger and the photosphere lies at a lower density.  The second feature just above 1~keV is caused by the vacuum resonance, which is where the effects of the plasma and vacuum birefringence are approximately equal.  In the figure, above the red line the plasma dominates the polarization modes and below the red line the vacuum dominates; it is clear then that all photons will pass through the vacuum resonance as they travel to us. However, low energy photons travel through the resonance unaffected, while photons of sufficiently high energy can swap modes through the resonance, i.e. photons in extraordinary mode below the resonance emerge in the ordinary mode above it. This effect brings the photosphere of the extraordinary mode closer to that of the ordinary mode above a few keV, but it still lies at higher density for the entire energy range considered.  As the temperature increases with density, the bulk of the radiation will emerge from the atmosphere in the extraordinary mode, and at the surface of the star, the radiation is more than ninety-five percent polarized in the extraordinary mode.

\begin{figure}
\centering
    \includegraphics[width=0.49\columnwidth]{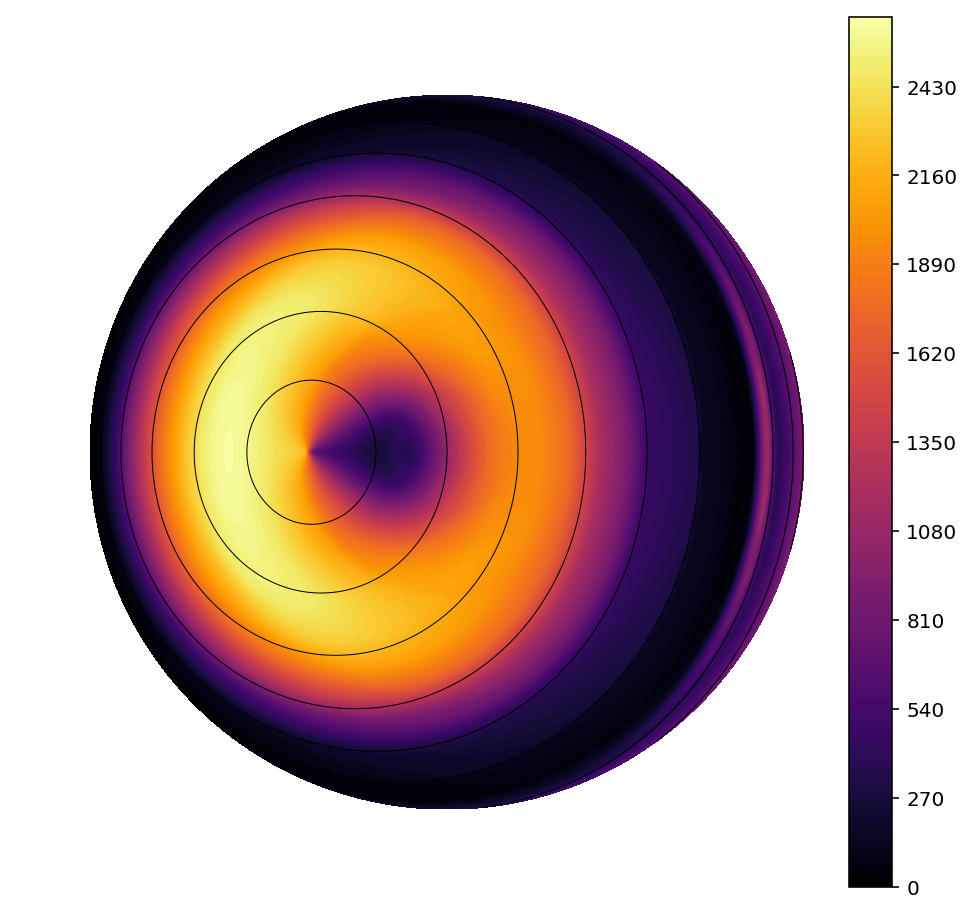} 
    \includegraphics[width=0.49\columnwidth]{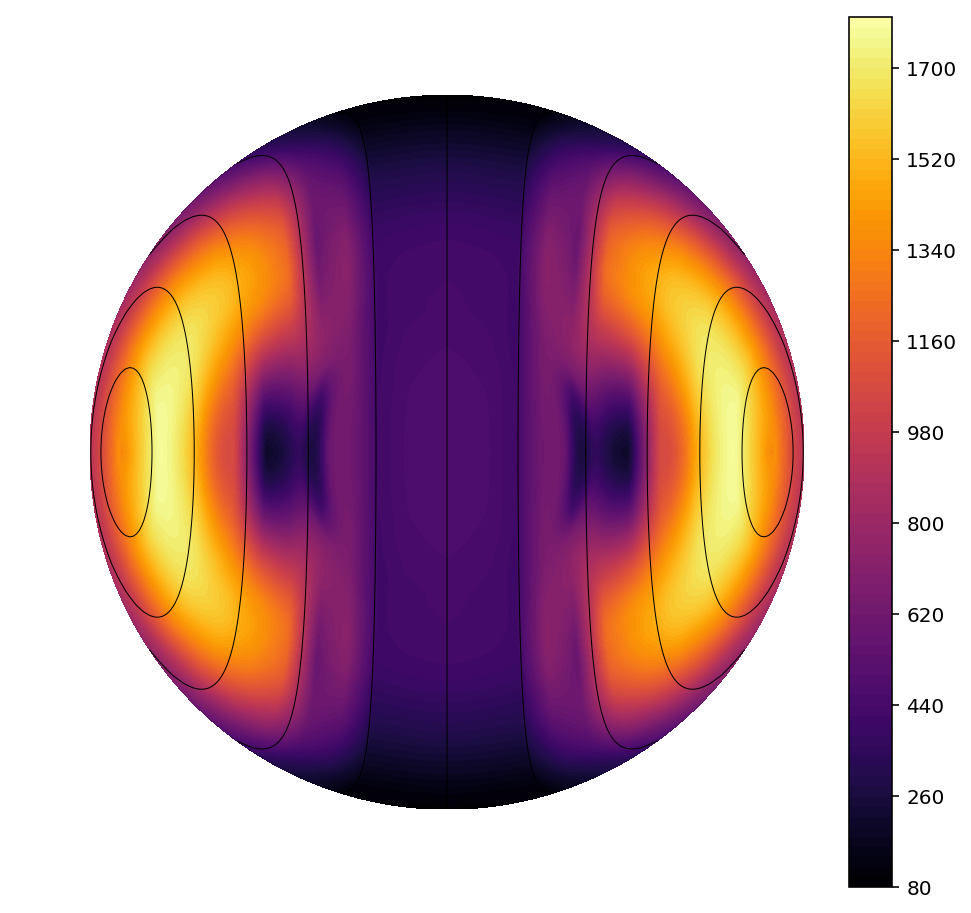} \\
    \caption{Total Emission from the Surface of a Magnetar.  The curves denote contours of constant magnetic latitude.  A dipole temperature map and fully ionized hydrogen atmosphere are included \citep{Caiazzo_2019}.  The colormap shows the intensity of the emission
for 2 keV photons in erg~cm$^{-2}$~s$^{-1}$~Hz$^{-1}$ sr$^{-1}$ in the observer's frame. }
    \label{fig:surfmap_total}
\end{figure}

The lower panel of Fig.~\ref{fig:atmo} depicts a second crucial consequence of the structure of the polarization states in the plasma: the radiation is highly beamed.  In fact, although radiation in the extraordinary mode emerges with a broad distribution in angles with respect to the magnetic field direction, the ordinary mode is strongly beamed along the magnetic field. 
As the direction of propagation of the photons approaches that of the magnetic field, neither mode has a large component along the direction of the field, so neither mode has a large cross-section.  Along the magnetic field lines, one can peer deep into the neutron star's atmosphere in both polarization states.  However, the beaming becomes very sharp and hard to resolve as the fields become stronger, so looking at the entire surface of the star as in Fig.~\ref{fig:surfmap_total} does not reveal the beaming along the field, but the diminished emission near the field direction is apparent.

If a polar hot spot is absent, the dipole emission pattern results in emission coming from the bulk of the stellar surface: the emission is brighter near the poles and nearly absent near the magnetic equator. However in this case the resulting pulsed fraction as the star rotates will be modest, as apparent from Fig.~\ref{fig:surfmap_total}, and gravitational lensing will reduce the pulsed fraction further \citep{1998MNRAS.300..599H,2016MNRAS.459.3585G}.  

\subsection{Condensed Surface}

Given the exotic conditions at the surface of a neutron star, the material at the surface may actually be condensed (solid or liquid) rather than gaseous \cite[e.g.][]{vanadelsberg05,medin07}.  Such a condensed surface means that the density drops abruptly, and the phase transition between the condensed phase and the vapour above marks the location of the photosphere of the star.  The star emits radiation as an incandescent metal.   This fact yields dramatic simplifications in the calculation of the emission.  One can assume that the surface is in equilibrium at a given temperature with the photons above it, and the calculation of the emission reduces to determining the coefficients of reflectivity and transmission between the photon modes in the diffuse plasma above the surface and the collective modes of the dense plasma below the surface.  Above the surface, the modes are again the ordinary and extraordinary modes as discussed earlier; at the surface itself, they are modes with the photon's electric field parallel to the surface ($s$) and those with a component normal to the surface $(p)$; below the surface, one has the full zoo of possibilities within a dense magnetized electron-ion plasma. 
\begin{figure}
\centering
\includegraphics[width=0.8\columnwidth]{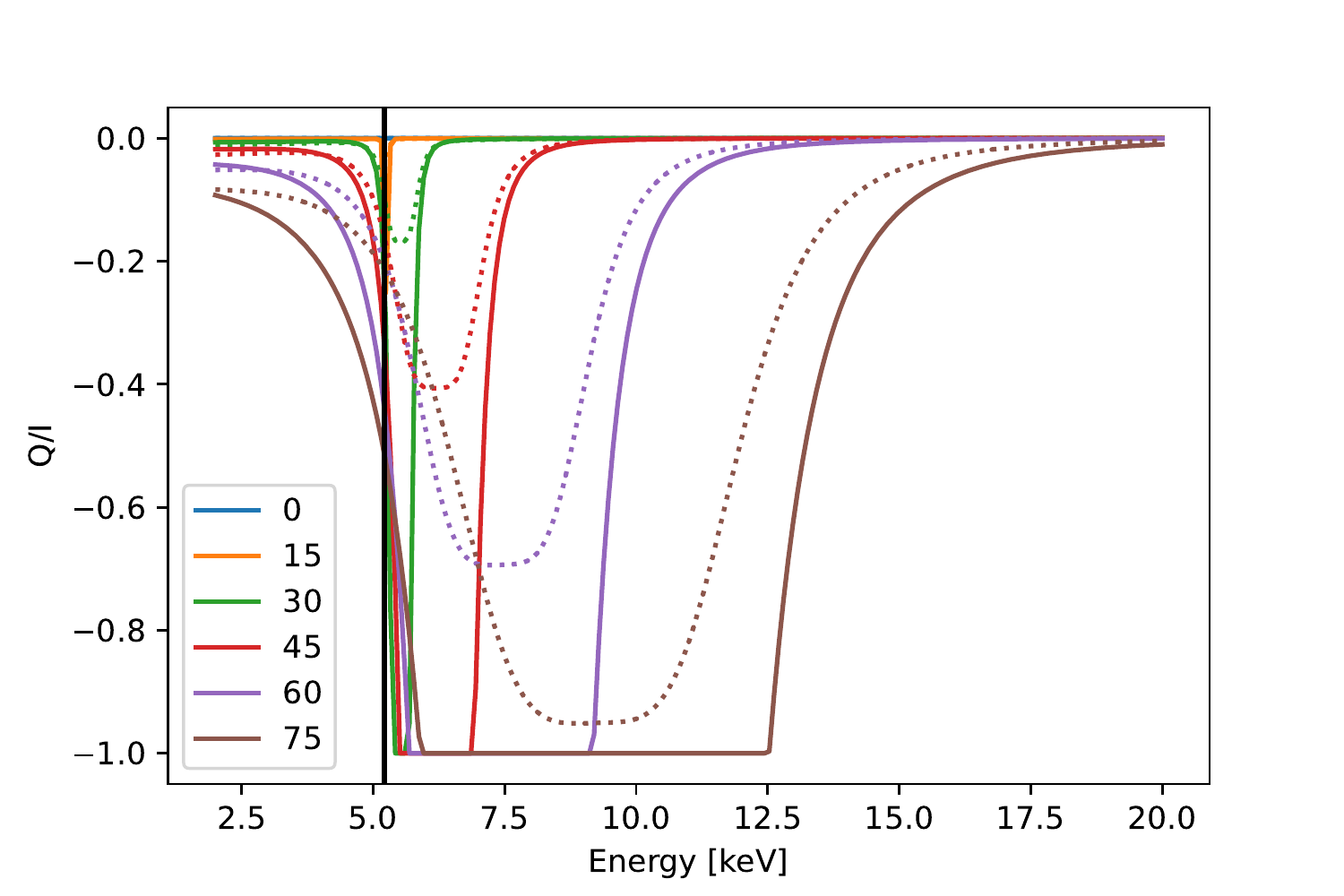}
\caption{The expected polarization from a condensed iron surface (solid lines for a solid surface, and dotted lines for a liquid surface) with a vertically directed magnetic field of $10^{14}$~G and various photon trajectories, zero through 75 degrees from the vertical \citep{2012A&A...546A.121P,2016MNRAS.459.3585G}.  A value of $Q/I$ equal to one corresponds to fully polarized in the ordinary mode.  The electron plasma frequency of 5.2~keV is denoted by a black vertical line.}
\label{fig:condensed}
\end{figure}

Fig.~\ref{fig:condensed} depicts the expected polarization from a condensed neutron star surface with a field of $10^{14}$~G.  The feature in the emissivity at about 5~keV results from the electron plasma frequency within the condensed iron surface.  Although this feature above the plasma frequency is quite pronounced for a single photon trajectory, \citet{2021arXiv211203401C} find that when one averages over the surface of the magnetar with a dipole emission pattern, this feature is much less apparent. The net polarization peaks at 15\% near the plasma frequency; otherwise it is smaller through the 2$-$10~keV range over which IXPE measures.  The key result here is that the extent of polarization from the condensed surface is much smaller than from a gaseous atmosphere, so IXPE observations should be able to determine whether the surface is condensed or not. On the other hand, the IXPE band is not conducive to probe whether the surface is liquid or solid.  In fact, observations in the visual band are best suited for this purpose: in magnetar fields, the solid state should emit mainly in the ordinary mode, while the liquid state also favours the ordinary mode but much less so, 10$-$30\% versus 70\%.  \citet{2016MNRAS.459.3585G} noted this effect in the more weakly magnetized and cooler XDINS and found a phase-averaged polarized fraction up to 30\% for the solid state.



\section{Scattering}
\label{sec:scattering}

We have seen that the action of vacuum birefringence in the magnetosphere of the neutron star will preserve the polarization state of the radiation from the surface to a large distance, if the radiation does not interact along the way with matter.  The spectral energy distributions of magnetars \citep[e.g.][]{2015ApJ...808...32T} indicate that the radiation does interact along the way.   In fact there are two additional components to the spectral energy distributions of magnetars: an excess from 4$-$15~keV and from 20$-$100~keV.   There are several models for both of these components, but we will focus on just resonant cyclotron scattering (RCS) for the high-energy excess \citep[e.g][]{fernandez07,nobili08}.  For the low-energy excess, which lies within reach of IXPE, we will look at three possibilities: a thermal polar hotspot, RCS, and complete Comptonization of photons in the ordinary mode.  In all three cases, our underlying emission model is a hydrogen atmosphere with dipole magnetic field and emission pattern.  Furthermore, we apply the effects of scattering in a semi-analytic fashion locally to the surface of the neutron star, using a rapid model prototyping package\footnote{https://github.com/UBC-Astrophysics/QEDSurface/}.   Although the models that we present here are restricted to \citet{lloyd03a} atmospheres with scattering, the package can also calculate emission from semi-analytic atmospheres \citep{1998MNRAS.298L..17H} and condensed surfaces \citep{2016MNRAS.459.3585G}.  For the high-energy component we simply add a power-law component with the necessary slope to account for the data and assign a polarization of $Q/I=-1/3$ as expected from the mode exchange cross-sections in RCS.

\begin{figure}
\centering
    \includegraphics[width=0.49\columnwidth]{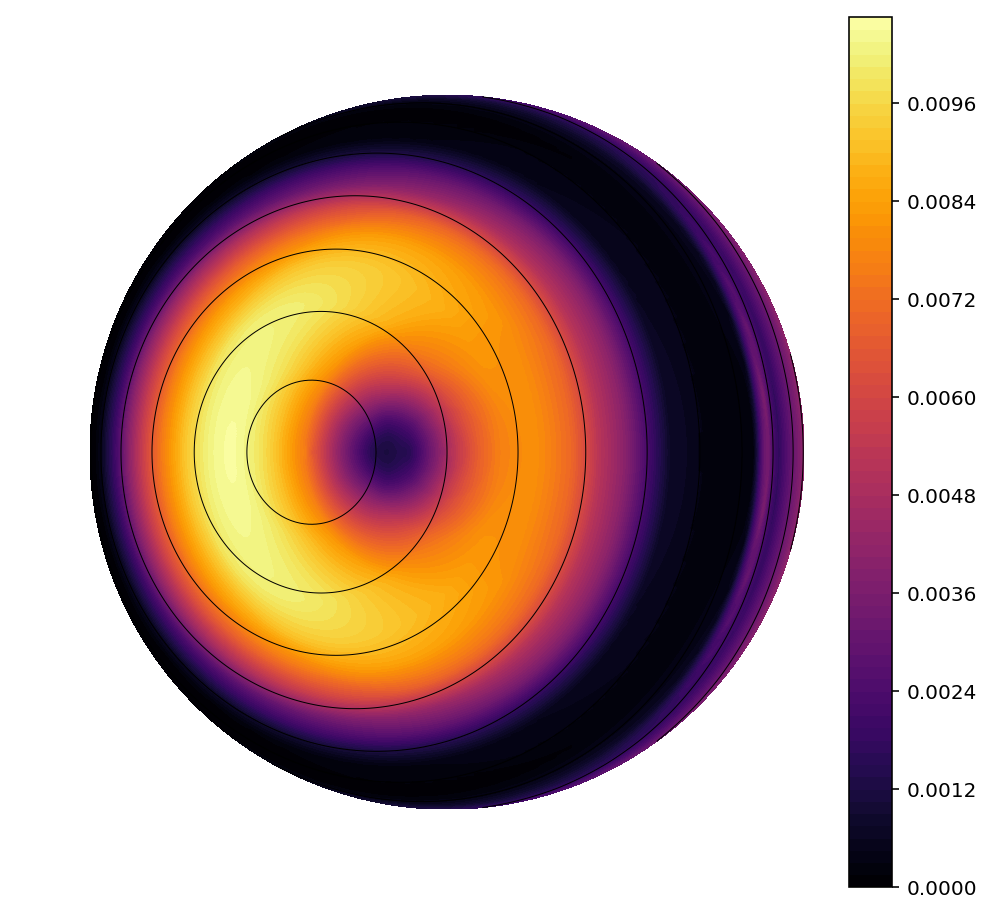} 
    \includegraphics[width=0.49\columnwidth]{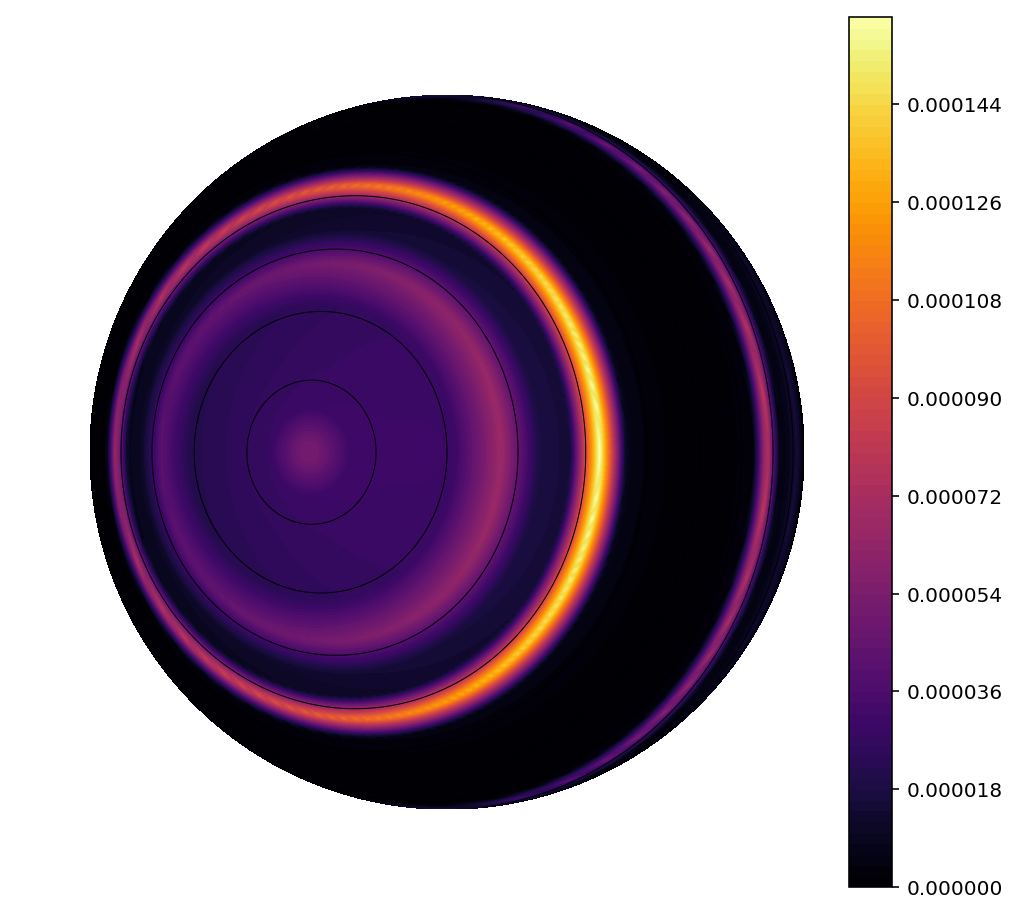} \\
\caption{Emission from the Surface of a Magnetar in the Extraordinary Mode (Left) and the Ordinary Mode (Right) including Compton scattering.  The curves denote contours of constant magnetic latitude \citep{Caiazzo_2019}. The colormap shows the intensity of the emission
for 5 keV photons in erg~cm$^{-2}$~s$^{-1}$~Hz$^{-1}$ sr$^{-1}$ in the observer's frame.}
\label{fig:omode}
\end{figure}

The polar hotspot model and the RCS model \citep{fernandez07,nobili08} are both familiar solutions to the excess from 4$-$15~keV \citep[e.g.]{2009MNRAS.398.1403Z} and yield definite predictions for the polarization, nearly completely extraordinary in the case of the hotspot and two-thirds extraordinary for RCS.  The ordinary mode scattering model is new \citep{Caiazzo_2019}.  Although the emission from a hydrogen atmosphere is nearly entirely in the extraordinary mode, a small portion is in the ordinary mode.  If one assumes that there is a diffuse layer of hot gas above the surface of the neutron star, radiation in the ordinary mode can be scattered into equilibrium with the electrons.  The required column density is just a several grams per square centimeter ($\tau$ of a few).  The scattering matrix well below the cyclotron energy almost always scatters ordinary mode photons into the ordinary mode and with a dipole-beam pattern perpendicular to the magnetic field.  Because the number of photons is preserved in this process, the final distribution is a Bose-Einstein distribution with the temperature of the electron gas and the same number of photons that as were in the ordinary mode initially. For both 4U~0142+62 \citep{2015ApJ...808...32T} and 1RXS J170849.0-400910 \citep{2009MNRAS.398.1403Z}, the number of ordinary mode photons produced by the atmosphere is sufficient to account for the excess flux from 4$-$15~keV observed in these sources.  Fig.~\ref{fig:omode} depicts what the surface of the magnetar 4U~0142+62  may look like in the extraordinary and the ordinary mode.   The ordinary mode appears concentrated in a ring at about magnetic co-latitude of sixty degrees.  Here the line of sight is approximately perpendicular to the local magnetic field, so the scattering preferentially directs the radiation toward the observer from these regions.

\section{Prospects}
\begin{figure}
\centering
\newcommand{\doplot}[1]{
\includegraphics[width=0.49\textwidth,clip,trim=0in 0.05in 0.6in 0]{figures/#1-TotalFlux.pdf}
\includegraphics[width=0.49\textwidth,clip,trim=0in 0.05in 0.6in 0]{figures/#1-PolFlux.pdf}
}
\doplot{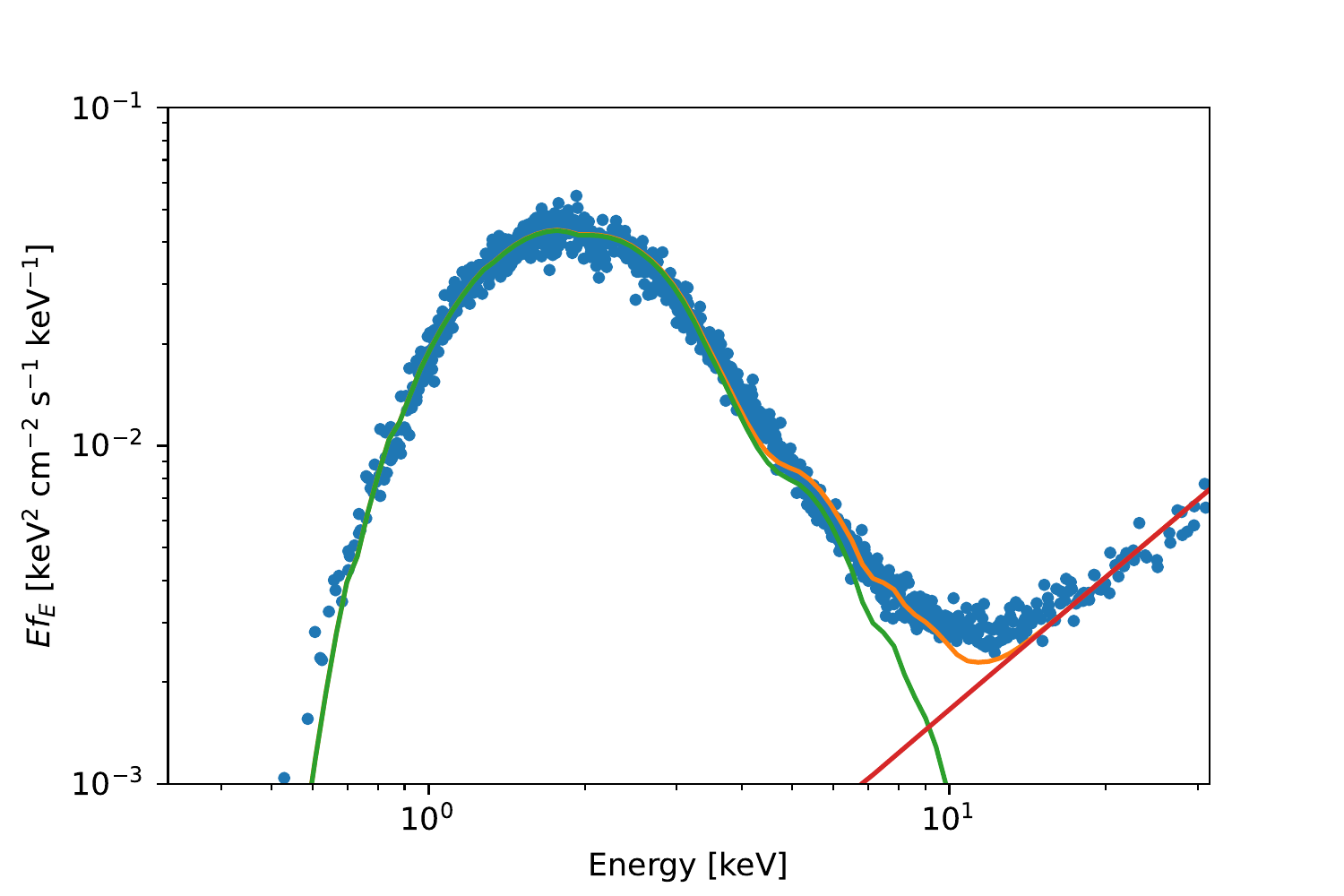}\\
\doplot{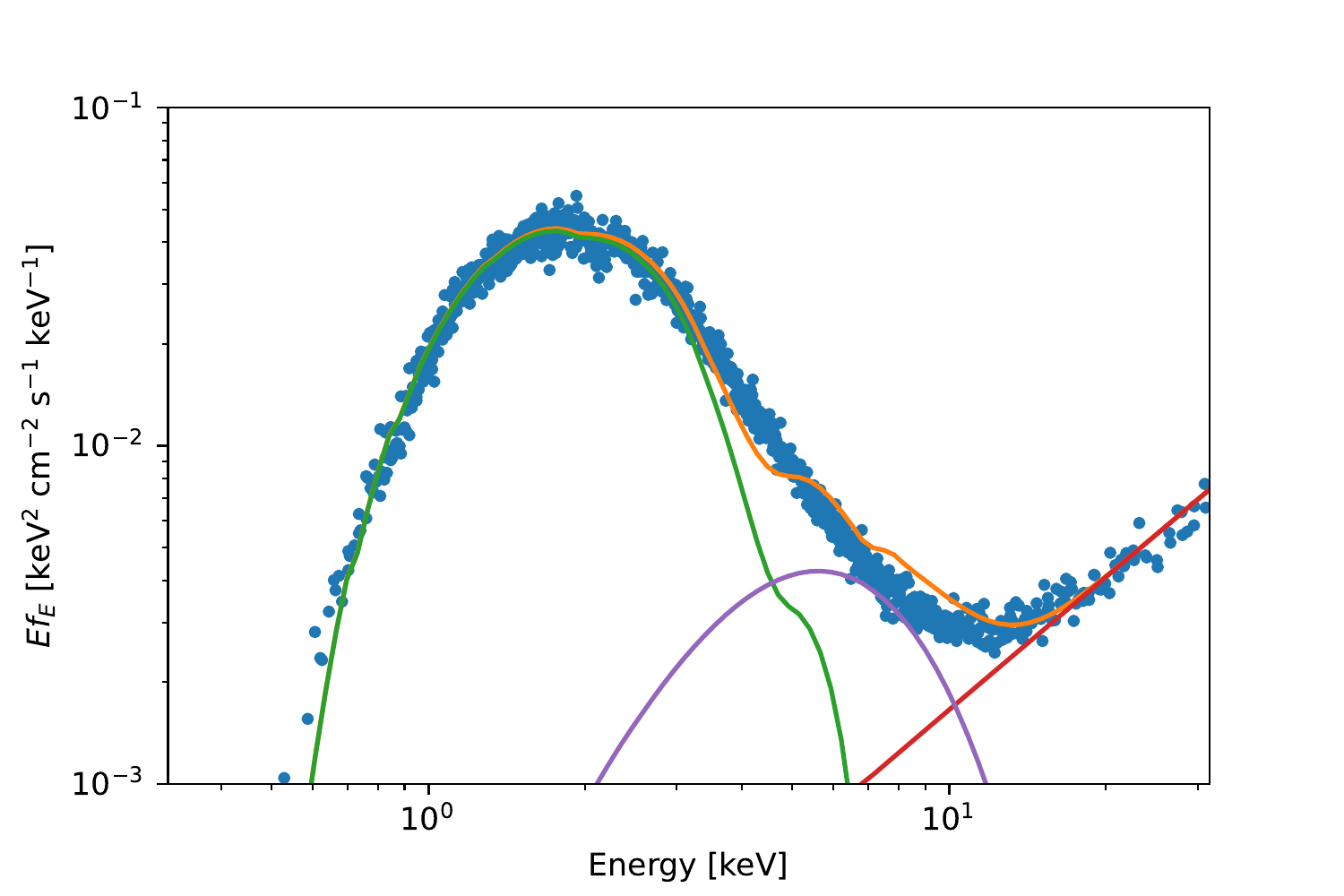}\\
\doplot{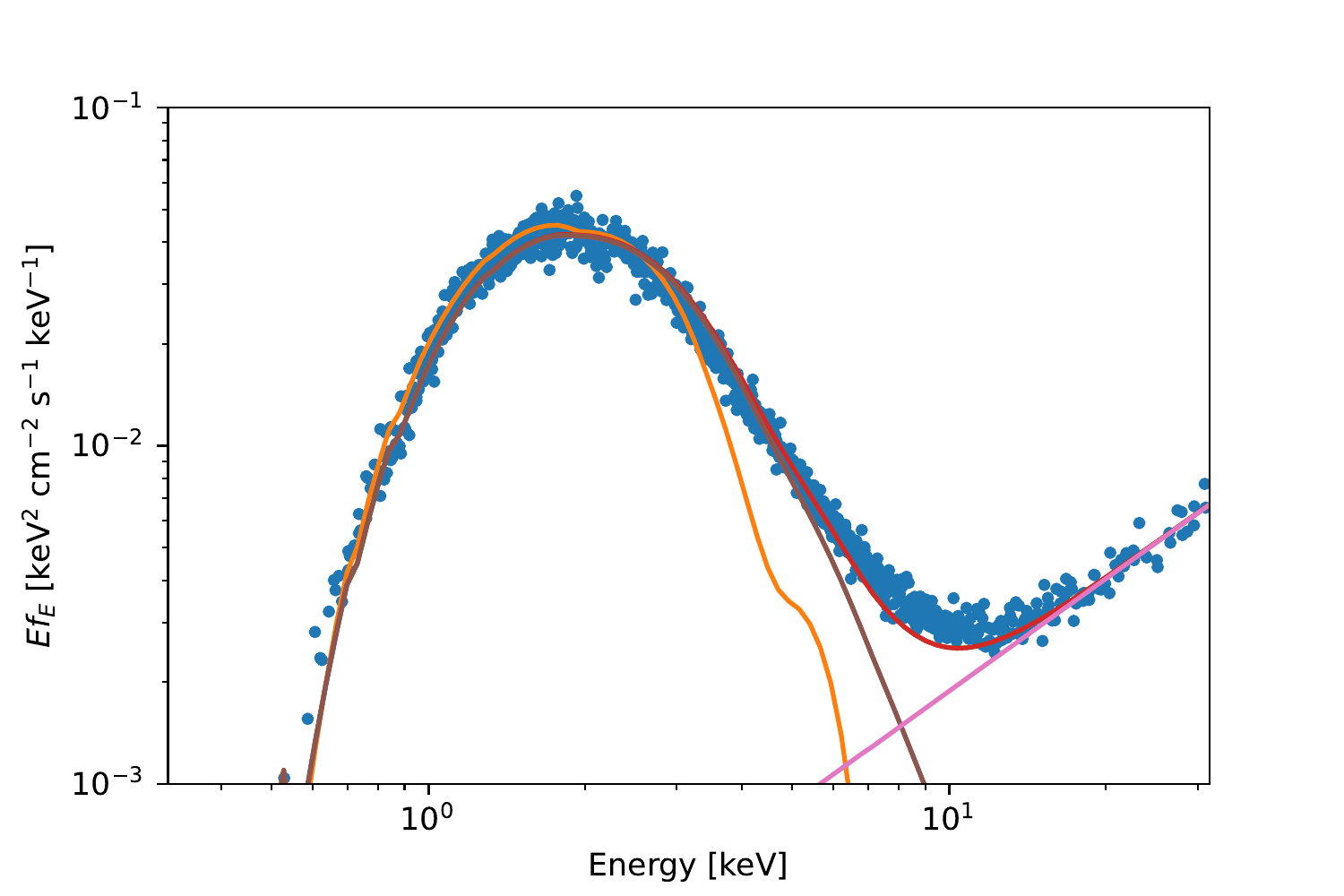}\\
\caption{Spectral shape and polarization for an atmospheric magnetar with different emission models for the ``soft excess''. The left panels show the spectral data of 4U 0142+61 \citep[blue dots,][]{2015ApJ...808...32T} and the different predicted spectral energy distributions. The right panels show the predicted polarization spectra; a $Q/I=-1$ indicates a 100\% polarization degree in X. \textbf{Upper panels: hot-spot model.} In the left panel, the different components are shown: the absorbed hydrogen atmosphere plus the hotspot (green) and the high-energy power-law, with spectral index $\Gamma=1.3$ (red). The orange line shows the sum of the 3 components and it is plotted on top of the spectral data for comparison. The right panel shows the same contributions but to the polarization spectrum; additionally, it shows the prediction for the same model when QED is neglected (blue solid line). \textbf{Middle panels: Comptonized O-mode model.} In the left panel, the different components are shown: the absorbed hydrogen atmosphere deprived of its O-mode photons (green), the Comptonized O-mode photons (purple) and the high-energy power-law, with spectral index $\Gamma=1.3$ (red). The orange line shows the sum of the 3 components and it is plotted on top of the spectral data for comparison. The right panel shows the same contributions but to the polarization spectrum; additionally, it shows the prediction for the same model when QED is neglected (blue solid line). \textbf{Lower panels: RCS model.} In the left panel, the different components are shown: the absorbed hydrogen atmosphere (orange), the RCS atmosphere (brown) and the high-energy power-law, with spectral index $\Gamma=1.3$ (purple). The red line shows the sum of the 3 components and it is plotted on top of the spectral data for comparison. In the right panel, the components in orange and brown are the same as for the left panel, for an electron temperature $kT_e=150$~keV. In purple RCS atmosphere and total (green) for an electron temperature $kT_e=100$~keV
\citep{Caiazzo_2019}}
\label{fig:magmodels}
\end{figure}

Observations of bright magnetars in quiescence with IXPE will provide a crucial test of the QED effect of vacuum birefringence as well as probe important properties of magnetars themselves; for example, these observations may reveal the state of their surfaces and the nature of their non-thermal emission.  Figure~\ref{fig:magmodels} shows how the polarization signal in the IXPE range will allow us to pinpoint the emission mechanism behind the ``soft excess'' detected in magnetar spectra between 4 and 15 keV. The left panels show the spectral energy distribution, while the right panels show the polarization signal predicted for a magnetar with an atmospheric surface under different models for the ``soft excess''. The three models can all reproduce the spectral shape of a persistent magnetar like 4U~0142+61, but they predict a polarization spectrum that differs significantly in the 1$-$10 keV range.  They can all account for the modest pulsed fraction of about 10\% for this source from 2-8~keV.

In the three simulations showed in the figure, the thermal emission is calculated using Lloyd's atmosphere code \citep[][see sec.~\ref{sec:atm}]{lloyd03a,lloyd03b} and imposing a surface temperature at the pole, $kT_{\textrm{eff}p} = 0.26$~keV, that can reproduce the thermal peak at $\sim 1$~keV, while the hard-energy tail is described by a simple power law with spectral index $\Gamma\sim1.3$. The three models employed for the ``soft excess'', described in section~\ref{sec:scattering}, are, from top to bottom: a thermal hot spot, the complete Comptonization of thermal O-mode photons by a corona of hot electrons and the RCS of atmospheric photons off mildly relativistic currents in the magnetosphere. The emission from the hot spot in the upper panel is calculated using Lloyd's code for a 2$^\circ$-diameter spot at the magnetic poles of the star with a temperature of $kT_{\textrm{eff}} = 1.06$~keV. In the fully Comptonized model (middle panel), all the O-mode thermal photons coming from the atmosphere are up-scattered by an electron-dominated plasma at a temperature $kT_{e} = 1.8$~keV. The RCS emission in the bottom panel is modelled assuming a partial Comptonization, with $y$ parameter $\sim0.1$, off a Maxwellian distribution of electrons at temperatures $kT_e=100$~keV and 150~keV. This is just a simple analytical toy-model that can reproduce the observed spectral shape; for a full RCS calculation see \citet{2021arXiv211203401C}.

The polarization degree is high in the X-mode at low energies for all cases, where the emission from the atmosphere dominates. For the high-energy tail, if one assumes a RCS model, the polarization is determined by the number of scatterings in the magnetosphere; the analysis of \citet{fernandez11} shows that the multiple scattering are important above $\sim10$~keV and that the polarization degree consequently converges to $\sim33\%$ in the X-mode due to mode exchange. The 3-10 keV range is where the models differ the most. In the hot spot case, where the dominant emission comes from the atmosphere even when the hot-spot dominates, the polarization is high and in the X-mode until the high-energy tail picks up. When the RCS is assumed to be also causing the ``soft excess'', in the lower panels, the transition toward the 33\%-level in polarization happens at lower energy. In the case of Comptonization, all the O-mode photons from the atmosphere are upscattered into the 4-15 keV range, where they dominate the emission, and therefore we see a switch between the X-mode and O-mode, which reaches a peak of 30\% polarized at 10 keV, before converging toward the 33\% X-mode in the high-energy tail.

The upper two rows of Figure~\ref{fig:magmodels} also show the predicted polarization signal when the QED effect of vacuum birefringence is neglected (solid light blue lines). As discussed in sec.~\ref{sec:QED}, the effect of QED is to comb and align the photons' polarization vectors, preserving the polarization at surface; when QED is neglected, the average over the different polarization directions on the surface of the neutron star yields a low polarization degree throughout the energy band. The only exception is for the hot-spot case: since we assumed a small sized spot, the magnetic field direction and therefore the polarization direction in the small region is reasonably aligned, and therefore when photons from the hotspot dominate the emission, we can see an increase in polarization fraction even when QED is not included (upper-right panel).

\section{Conclusion}

Observations with IXPE will provide unprecedented probes of the properties of magnetars and their phenomenology.  We have outlined how IXPE observations of the AXP 4U~0142+61 will measure the effects of vacuum birefringence definitively, probe the physical state of the surface of the magnetar and the role of currents in the magnetosphere of the neutron stars.  These observations will provide verification of the magnetar model and QED as well as important clues to the origin of the high-energy emission from magnetars.

\bibliographystyle{aasjournal}
\bibliography{other}
\end{document}